\documentclass[review]{elsarticle}

\usepackage{hyperref}

\journal{Journal of \LaTeX\ Templates}

\usepackage{graphicx,txfonts}

\def\simgreat{\mathbin{\lower 3pt\hbox
     {$\rlap{\raise 5pt\hbox{$\char'076$}}\mathchar"7218$}}}

\def\simless{\mathbin{\lower 3pt\hbox
     {$\rlap{\raise 5pt\hbox{$\char'074$}}\mathchar"7218$}}}

\newcommand{\ga}{\ge}
\newcommand{\la}{\le}
\newcommand{\arcsec}{$''$}
\newcommand{\degr}{^{\rm {o}}}

\begin{document}

\begin{frontmatter}

\title{Constraints on interstellar dust models from extinction and spectro-polarimetry}

\author {R.~Siebenmorgen}

\address{European Southern Observatory, Karl-Schwarzschild-Str. 2, D-85748 Garching b. M\"unchen, Germany, Ralf.Siebenmorgen@eso.org}

\author {N.V. Voshchinnikov}
\address{Sobolev Astronomical Institute, \\ 
      St.~Petersburg University, Universitetskii prosp. 28, St.~Petersburg, 198504 Russia }
\author {S. Bagnulo}
\address{Armagh Observator and Planetariumy, College Hill, Armagh BT61 9DG, UK}

\author {N.L.J.~Cox} \address{ Anton Pannekoek Institute for
  Astronomy, University of Amsterdam, NL-1090 GE Amsterdam, The
  Netherlands.}

\begin{abstract}

  We present polarisation spectra of seven stars in the lines-of-sight
  towards the Sco OB1 association. Our spectra were obtained within
  the framework of the Large Interstellar Polarization Survey carried
  out with the FORS instrument of the ESO VLT.  We have modelled the
  wavelength-dependence of extinction and linear polarisation with a
  dust model for the diffuse interstellar medium which consists of a
  mixture of particles with size ranging from the molecular domain of
  0.5\,nm up to 350\,nm. We have included stochastically heated small
  dust grains with radii between 0.5 and 6\,nm made of graphite and
  silicate, as well as polycyclic aromatic hydrocarbon molecules
  (PAHs), and we have assumed that larger particles are prolate
  spheroids made of amorphous carbon and silicate.{{ Overall, a
      dust model with eight free parameters best reproduces the
      observations. Reducing the number of free parameters leads to
      results that are inconsistent with cosmic abundance
      constraints.}} We found that aligned silicates are the dominant
  contributor to the observed polarisation, and that the polarisation
  spectra are best-fit by a lower limit of the equivolume sphere
  radius of aligned grains of 70 -- 200\,nm.

\end{abstract}

\begin{keyword}
dust, extinction, polarisation, interstellar medium
\end{keyword}

\end{frontmatter}

%\linenumbers

\section{Introduction}

Our understanding of many astrophysical processes, ranging from galaxy
evolution to stellar and planetary formation, depends on our knowledge
of the cosmic dust (e.g. \citet{Draine09, Asano14}). Dust and
molecular gas co-exist in interstellar clouds, and the properties of
both components are mutually dependent through various chemical
reactions and physical (photon) processes.

Interstellar dust grains absorb, scatter, and polarise the background
radiation. The observed wavelength dependencies of extinction,
emission and polarisation of the radiation coming from background
sources may allow us to characterise the dust of the diffuse
interstellar medium (ISM) (\citet{Desert, S92, Kim, Dwek,
  Weingartner, Draine07, DF09, Jones13, K15, V12, S14, V15}).

The absolute extinction is often represented as a function
$A(\lambda^{-1})$ normalised by the extinction value in the visible
band $A_V$.  This extinction curve $A(\lambda^{-1})$ has been {
  {characterised}} for hundreds of sight lines (\citet{FM86, FM90,
  Calzetti94, Papaj, FM07, Valencic, Gordon}).  It varies with
wavelength and approaches zero at longer wavelengths.  In the optical,
the extinction curves tend to follow a relation that depends on the
total-to-selective extinction $R_{{V}} = A_{{V}} / E_{{B - V}}$
(\citet{Cardelli88}, see also \citet{FM07}), where $E_{{B - V}} = (B -
V) - (B - V)_0$ is the colour excess given by the observed magnitude
difference at $B$ and $V$ of the star and of identical unreddened star
(\citet{Johnson, Bless70, Massa, Massa90}).  Flat extinction curves
with large $R_{{V}}$ values are measured towards denser regions. The
parameter $R_{{V}}$ is also a rough indicator of grain size: sight
lines of low $R_{{V}}$ values are thought to be characterised by
smaller grains than sight lines with higher $R_{{V}}$ values.

Extinction curves may be modified by scattering of photons from dust
clouds in or out of the line of sight.  \citet{Scicluna15} have shown
that the observed extinction is not significantly modified by
scattering from dust clouds at distance smaller than 1\,kpc. For such
nearby stars, the extinction curves provide crucial information on the
composition and size distribution of the interstellar dust.

Also polarisation spectra constrain sizes, shape and composition of
the dust grains. Asymmetrical dust grains are most likely aligned with
the magnetic field and polarise the light (\citet{Andersson15}). So
far, the majority of polarisation measurements are obtained in optical
($UBVR$) and near infrared ($IJHK$) broadband filters
(\citet{Whittet}).  The observed polarisation spectra generally have a
quasi-parabolic shape (\citet{Serkowski}). The position of the
polarisation maximum and the width of the polarisation spectrum varies
from star to star, and this diversity is probably due to different
local conditions affecting the alignment efficiencies and the grain
sizes (\citet{V14}).  Insight into the shape and size distribution and
composition of interstellar grains may come from detailed physical
models (\citet{DF09, V12, S14, V15}).

We have started a Large Interstellar Polarisation Survey (LIPS; PI:
N.J.L.~Cox) with the aim of determining the chemical composition and
size distribution of interstellar dust in numerous sight-lines.  In
this paper we present spectro-polarimetry of seven early-type stars
obtained with the FORS2 instrument (\citet{Appenzeller98}) of the ESO
VLT.  Spectra cover the wavelength range 365 -- 920\,nm and have a
resolution of about 880.  For our sample of stars, extinction curve
measurements are available by \citet{Valencic}. The seven stars are
located towards sight lines of the Scorpius--Centaurus Association
(called Sco OB1).

This paper is organised as follows. In Sect.~\ref{obs.sec} we present
our new data. In Sect.~\ref{model.sec} we present a dust model and a
procedure for fitting the extinction and polarisation curve of an
object simultaneously.  Sect.~\ref{result.sec} presents the result of
our modeling efforts. In Sect.~\ref{conclusion.sec} we give a summary
of our main results.

\section{Observations} \label{obs.sec}

From the original LIPS target list we selected a small sub-sample
comprising seven close-by lines-of-sight towards the Sco OB1 region,
including two possible association members, HD\,151804, HD\,152235
that are supergiants stars.

\begin{figure}
\centering
\includegraphics[width=12cm,clip=true,trim=0.cm 0cm 0.cm 0cm]{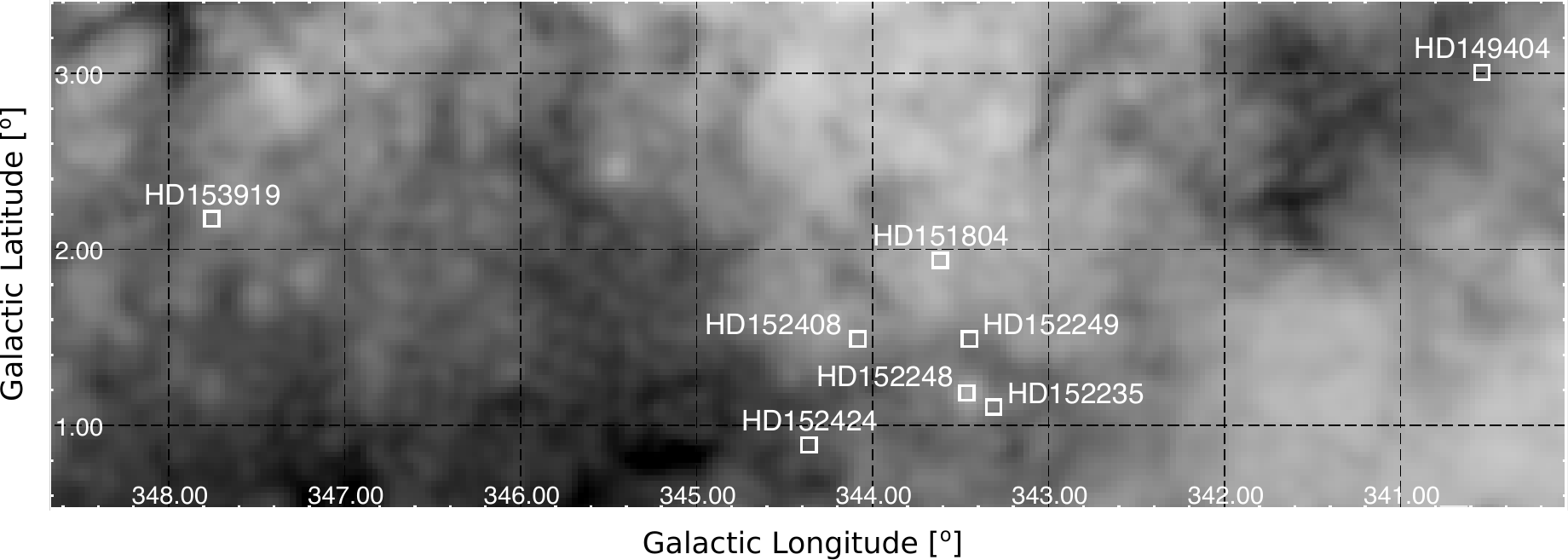}
\caption{Observed stars towards Sco OB1 displayed on top of an
  extinction map produced from DSS
  (\citet{Dobashi}). \label{targetmap.fig}}
\end{figure}

In Fig.\ref{targetmap.fig} we show the target location in galactic
coordinates overlaid to an extinction map produced from DSS
(\citet{Dobashi}).  The distance for HD\,153919 is 1.7\, kpc
(\citet{Ankay}) while the other stars (HD\,151804, HD\,152235,
HD\,152248, HD\,152408), are confirmed Hipparcos members of Sco OB1
which is at an estimated distance of 2\,kpc. The stars HD\,152235,
HD\,152248, HD\,152249 and HD\,152408, are close to each other in the
sky, and have low polarisation efficiency of $p_{\rm {max}}/E(B-V)
\sim 1-2$.  More detailed target information is given in
Table~\ref{target.tab}.

\begin{table*}
  \caption{Observing sample. } \label{target.tab} 
\begin{center}
\begin{tabular}{lllllll}
\hline\hline
  Target	 & Sp.Type    & $l$	 & $b$	    & V    & E(B-V) 	     & R(V)	 \\
 (1) & (2) & (3) & (4) & (5) & (6) & (7) \\
\hline
  HD\,149404 & O8.5Iab(f)p& 340.5375 & +03.0058 & 5.87 & 0.62$\pm$0.06   & 3.53$\pm$0.38 \\
  HD\,151804 & O8Iaf      & 343.6156 & +01.9378 & 5.29 & 0.30$\pm$0.03   & 4.33$\pm$0.30  \\
  HD\,152235 & B0.5Ia     & 343.3111 & +01.1041 & 6.38 & 0.71$\pm$0.06   & 3.13$\pm$0.25  \\
  HD\,152248 & O7Iabf+O7Ib& 343.4644 & +01.1839 & 6.15 & 0.41$\pm$0.04   & 3.68$\pm$0.26 \\
  HD\,152249 & OC9Iab     & 343.4488 & +01.1649 & 6.65 & 0.46$\pm$0.10   & 3.54$\pm$0.45 \\
  HD\,152408 & O8Iape     & 344.0836 & +01.4909 & 5.92 & 0.42$\pm$0.05   & 4.17$\pm$0.33 \\
  HD\,153919 & O6Iafcp    & 347.7544 & +02.1735 & 6.78 & 0.51$\pm$0.04   & 3.87$\pm$0.22  \\
\hline
\end{tabular}
\end{center}

{{NOTES} -- Target name (1), Spectral type of star (2), galactic
  longitude $l$ (3) and latitude $b$ (4), $V$ from Simbad (5),
  $E(B-V)$ (6) and $R(V)$ (7) by \citet{Valencic}. }
\end{table*}

\subsection{Extinction}

Extinction curves are usually measured with ground-based facilities in
the near IR, optical and UV photometric bands, and, at shorter UV
wavelengths, by the IUE ($3.3\,\mu\rm{m}^{-1} < \lambda^{-1} <
8.6\,\mu\rm{m}^{-1}$) and FUSE ($3.3\,\mu\rm{m}^{-1} < \lambda^{-1} <
11\,\mu\rm{m}^{-1}$) satellites, which provide spectroscopy at a
resolution of $\sim 0.5$\,nm. Extinction data are often fit by a third
order polynomial and a Drude profile to account for the 217\,nm
extinction bump (\citet{FM90, FM07}). Extinction curves of our sample
stars are provided by \citet{Valencic} using IUE data, while FUSE
observations at $\lambda^{-1} < 8.6\,\mu \rm{m}^{-1}$ are not
available for our stars.  Following \citet{Gordon}, we decreased the
value of the parameter describing the far UV rise as given by
\citet{Valencic} by $\sim 8$\%; other fit parameters determined with
IUE data, or those obtained using the combined IUE and FUSE data sets,
are broadly consistent (\citet{Gordon}). In order to fit the
extinction data in the $UBVJHK$ bands we have used the $R_{{V}}$
parametrisation by \citet{Fitzpatrick04}.

Extinction curves of the ISM that are derived from observations of
supergiants might be affected by systematic errors that are larger
than for main sequence stars. For supergiants it is difficult to
disentangle the ISM extinction from contribution by circumstellar dust,
and it is difficult in finding an unreddened supergiant with spectral
class. The extinction curves of our sample stars have typical
uncertainties of 10\% in the optical/UV with somewhat larger
uncertainties in the near infrared.

\subsection{Polarimetry with FORS2}
Linear spectro-polarimetric observations were obtained with the FORS2
instrument of the ESO VLT (\citet{Appenzeller98}) within the context
of the Large Interstellar Polarisation Survey (LIPS, PI:
N.J.L. Cox). All observations were obtained with grism 300V, which
covers the spectral range $365 - 920$\,nm, using the so-called
beam-swapping technique recommended in the FORS user manual, and
described in detail, e.g., by \citet{Bagnulo09}. With a
0.5\arcsec\ slit width, our observations have a nominal spectral
resolution $\lambda/\Delta \lambda \sim 880$. Data were reduced
following the guidelines by \citet{Bagnulo09}. Note that null profiles
could not be determined since observations were obtained with four
exposures only (at the following four positions of the retarder
wave-plate: $0^\circ$, $22.5^\circ$, $45^\circ$, $67.5^\circ$). To
improve the S/N ratio, we decided to rebin our spectra, by degrading
the spectral resolution to $\lambda/\Delta \lambda \sim 50$.  In the
rebinning process, outliers (e.g., due to cosmic-rays) were rejected
with a 3\,$\sigma$ clipping algorithm; after outlier rejection, we
calculated the weighted mean of the pixel values, and estimating
1\,$\sigma$ error from the variance of the polarisation of the
original bin. In Table~\ref{obs.tab} we specify the observing date and
time, the on source exposure time, and the mean signal-to-noise per
\AA.

%%%%%%%%%%%%%%%%%%%%%

\begin{table*}
  \caption{FORS/VLT spectro-polarimetric observation log.
Columns 2 and 3 refer to the date and time of the
observations, col.~4 is the exposure time in secs, and col.~5
provides the S/N per \AA\ at peak level.  } \label{obs.tab} 
\begin{center}
\begin{tabular}{ccccc}
\hline\hline
Star        &   Date        &   UT    & EXP  & S/N     \\
            &  (yyyy-mm-dd) & (hh:mm) & (s)  & (per \AA\,) \\
  (1)       &  (2)          &   (3)   &  (4) & (5) \\
\hline
%HD\,111934 & 2015-04-07 & 08:57 & 6.0  &  85 \\
HD\,149404 & 2015-05-12 & 05:40 & 1.6  &  80 \\
HD\,151804 & 2015-05-24 & 03:41 & 1.6  &  75 \\
HD\,152335 & 2015-05-24 & 04:01 & 2.0  &  70 \\
HD\,152248 & 2015-05-24 & 04:18 & 2.0  &  75 \\
HD\,152249 & 2015-05-24 & 04:34 & 4.0  &  80 \\
HD\,152408 & 2015-05-24 & 04:49 & 1.6  &  80 \\
%HD\,152424 & 2015-05-24 & 05:07 & 2.0  &  65 \\
HD\,153919 & 2015-04-12 & 09:01 & 4.0  &  80 \\
\hline
\end{tabular}
\end{center}
\end{table*}

\section{Model} \label{model.sec}
In this section we describe our technique for simultaneous modelling
of UV to IR dust extinction and linear polarisation curves observed
towards the diffuse ISM.  

\subsection{Dust populations} \label{dustpop.sec}
Data cannot be fit by assuming that all dust particle have the same
size, but we need to introduce a dust grain size distribution. In
particular we use a power law size distribution $n(r) \propto r^{-q}$
as introduced by \citet{Mathis77}. Particle sizes range from the
molecular ($r_{-} \sim 0.5$\,nm) to the sub-micrometer domain.

Interstellar polarisation cannot be explained by spherical particles
made of optically isotropic material, but we need to assume that some
dust particles are of non--spherical shape, and partly
aligned. Therefore we have considered spheroidal dust particles.  In
the following we will denote with $a$ and $b$ their major and minor
semi-axis, respectively. There exist two types of spheroids: oblates
(or 'disks') and prolates (or 'needles').  Oblate spheroids are
obtained by a rotation of an ellipse around the minor ($b$) axis and
prolates are obtained by a rotation of an ellipse around the major
($a$) axis. The sizes of spheroids can be compared if we use the
sphere of radius $r$ which volume equal to that of spheroid:
$r^3=ab^2$ for prolate spheroids and $r^3=a^2b$ for oblate ones. We
found that prolate particles provide better fits to the polarisation
spectra than oblate particles.

\citet{V12, Das10, S14} computed extinction ${C}_{\rm ext}(\nu)$,
scattering ${C}_{\rm sca}(\nu)$, and linear polarisation $ {C}_{\rm
  p}(\nu)$ cross sections and corresponding efficiency factors $Q =
C/\pi r^2$ using the separation of variables method presented by
\citet{V93}. Subsequently, cross sections were averaged over size
distribution and rotations.

In our dust model we include large homogeneous {\it spheroids} that
are made up of silicate and amorphous carbon, and a population of
small graphite and small silicates, as well as a PAH component with 60
H atoms and 150 C atoms, with cross sections ${C}_{\rm PAH}(\nu)$ as
given by \citet{S14}. Dust cross-sections of prolate particles are
computed for 100 radius values between 6 and 800\,nm.  Small grains
have radius $r$ such that 0.5\,nm $\le r \la 6$\,nm. \citet{Draine16}
pointed out that the rapid fall-off in starlight polarisation in the
far UV suggests that grains in this size range are either nearly
spherical, or do not contribute to the far UV extinction.  In our
model they contribute to the far UV extinction and do not contribute
to the polarisation curve. Therefore we assume that grains below 6\,nm
have spherical shape.  We apply for large silicates and carbon
particles different upper size limits $r_{+}^{\rm {Si}}$ and
$r_{+}^{\rm {aC}}$.

For graphite, the dielectric function is anisotropic and average
extinction efficiencies are computed by setting ${Q} = 2
{Q}(\epsilon^\bot)/3 + {Q}(\epsilon^\Vert)/3 $, where $\epsilon^\bot$,
$\epsilon^\Vert$ are dielectric constants for two orientations of the
$\vec{E}$ vector relative to the basal plane of graphite
(\citet{Draine93}). Efficiencies in both directions are computed via
Mie theory. Optical constants for large silicates and graphite are
given by \citet{Draine03}. { {Hydrogen-free amorphous carbon
    becomes hydrogenated, when exposed to atomic hydrogen
    (\citet{Furton93}). We employ the ACH2 mixture by
    \citet{Zubko}. This are hydrogenated amorphous carbon particles
    and semiconductors.  Densities of these grains depend on their
    sp2/sp3 ratio and hydrogen content, and range from 1.2 -- 2.2 \,
    (g/cm$^3$) (\citet{Robertson02, Casiraghi05}).  \citet{Furton99}
    characterized the interstellar absorption feature at 3.4\,$\mu$m
    by such grains and estimate a mass density of 1.5\,g/cm$^3$ for
    them.  }} We take a bulk density $\rho_C \sim 1.6$\,g\,cm$^{-3}$
for carbon and $\rho_{\rm Si} \sim 3.5$\,g\,cm$^{-3}$ for
silicates. We also computed dust cross sections with optical constants
provided by \citet{Jones12} and \citet{Jones13}. We find that in the
optical/UV the dust cross sections are similar for either set of
constants, whereas differences in the cross sections become strong at
wavelengths $\ga 1\,\mu$m. In summary, four different dust populations
are considered and in the following labelled as large silicates (Si),
large amorphous carbon (aC), small silicates (sSi), small graphite
(gr) and PAH. 
     
We fit the extinction curve, or equivalently the observed optical
depth profile, for each star of our sample by the extinction cross
section of the dust model. The total dust extinction cross section
$K_{\rm ext}({\nu})$ is given in cm$^2$\,g$^{-1}$. It is wavelength
dependent and given as the sum of the absorption and scattering
coefficient of the dust materials. For each dust component there is a
specific (relative) dust abundance. It is denoted by $\Upsilon$
together with a subscript for each dust population ${\rm {Si, aC, sSi,
    gr, PAH}}$. The specific dust abundance avoids introducing
systematic errors of present estimates of the gas phase depletion of
an element (\citet{Draine11, Dwek16, Gerin, Nieva, Jenkins, Parvathi,
  VH10}). Formulas for computing $K_{\rm {ext}}$ and $\Upsilon$ are
given in \citet{S14}.

\subsection{Linear polarisation}
The extinction cross section of non--spherical particles depends on
their orientation with respect to the electric field of the incident
light.  Unpolarised stellar light that passes a cloud of moderate
extinction by aligned grains becomes linearly polarised.  In the
spectral region from the far UV to the near IR, the observed
polarisation curves may be fit by the empirical formula given by
\citet{Serkowski}

\begin{equation}
{p(\lambda)} = {p_{\max}} \ \exp \left[ -k_{\rm p} \ \ln^2
    \left( \frac{\lambda_{\max}}{\lambda} \right) \right]\,.
\label{serk.eq}
\end{equation}

Equation~(\ref{serk.eq}) includes three parameters: the maximum
polarisation $p_{\max}$, the wavelength $\lambda_{\max}$ at
$p_{\max}$, and the width of the spectrum $k_{\rm p}$. \citet{V15}
studied observations of 160 sight lines of mildly reddened stars. They
report that the observed linear polarisation is of a few percent, $p
/E_{B-V} \la 9$\,\%/mag and its maximum $p_{\rm {max}}$ is always $<
10$\,\%. The width of the polarisation curve is $0.5 \la k_{\rm p} \la
1.5$, and $\lambda_{\rm max}$ is often observed close to the $V$
band. It varies between $0.35 - 0.8\,\mu$m and exceptionally up to $\sim
1\,\mu$m in dark clouds (\citet{Goodman}).

Observed values of the Serkowski parameters can be modeled assuming
aligned and homogeneous spheroidal sub-micrometer sized dust particles
made of silicates and carbon.  One simplifies the alignment process
assuming picket fence (\citet{K08, DF09}), the radiative torque (RAT,
\citet{Lazarian03,Andersson15}) or the imperfect Davies Greenstein
(IDG) alignment (\citet{V12}).  For the diffuse medium, the IDG
function is often applied. It is analytic and depends on particle
size, dust precession angle, the dust-to-gas temperature ratio $T_{\rm
  d}/T_{\rm g}$, and an alignment efficiency parameter $\delta_0$.
The maximum polarisation depends strongly on $\delta_0$ and on the
magnetic field orientation $\Omega$, which is the angle between $B$
and the line of sight.  The two quantities are difficult to be
disentangled from each other.  Fortunately, they have little influence
on the normalised polarisation curve, which is parameterised by
$\lambda_{\rm max}$ and $k_{\rm p}$.

The influence of dust parameters on polarisation was studied by
\citet{Das10, V14, S14, V15}, who considered a dust size distribution
with the equivolume sphere radius ranging from $r^{\rm pol}_-$ up to
$r^+$. They found that oblate spheroids produce narrow polarisation
curves often outside the observed values of $\lambda_{\rm max}$ and
$k_{\rm p}$, and that they are more efficient polarisers than prolates
(hence they may explain larger values of $p_{\max}$). The aspect ratio
$a/b$ of the spheroids has no strong impact on the Serkowski curve.
Adopting prolate grains generally leads to better fits to polarisation
data. As the ratio $a/b$ between the semi-axis of the prolates
increases, the Serkowski curve widens and the $\lambda_{\rm max}$
value decreases. The exponent $q$ of the size distribution of prolates
does not influence the polarisation strongly. The model parameters
$r_{+}$ and more so $r_{-}^{\rm {pol}}$ are those with the major
impact on the polarisation curve.

In this work, we model the observed linear polarisation using

\begin{equation}
  {p(\nu) \over A_{V}} = {K_{\rm {p}}(\nu) \over K_{{\rm {ext},} V}}\,.
\label{ptau.eq}
\end{equation}

where $K_{\rm p}(\nu)$ is the linear polarisation cross section of the
dust (\citet{V12, S14}).

\subsection{Fitting procedure}
We computed the various dust cross-sections in the spectral range from
70\,nm up to 2\,mm and for grain radius between 0.5\,nm and
800\,nm. We defined a grid size by setting $r_{i+1} = 1.05 \ r_i$, and
we linearly interpolated the modelling results for the size values
intermediate between two consecutive grid points. We considered a
different upper grain radius for silicates and carbon.  The other free
parameters of our models are the exponent of the dust size
distribution and the specific dust abundances.

We modelled the shape of the extinction and polarisation curves and
not their absolute properties, therefore it was sufficient to specify
for the $n=5$ dust populations only $n-1$ {{specific}} dust
abundances $\Upsilon$.  The abundances could be scaled up or down by
an arbitrary factor without any change in the predicted extinction or
polarisation curves. We experimented with a normalisation of the
specific dust abundances $\Upsilon$ that allows a direct comparison
and consistency check with absolute abundance constraints.  We set
$\Upsilon_{\rm {Si}}= 15$\,ppm in large silicate grains, { {which
    equals the lower limit derived by \citet{VH10}.}}

We modeled the shape of the polarisation curve but not its absolute
value $p_{\max}$, which allows to keep the IDG alignment parameters
and the field orientation constant. We use $T_{\rm {d}} /T_{\rm {g}}
=0.1$, $\delta_0 =10\mu$m, and $\Omega =90^{\degr}$ (for a study of
the influence of these parameters on the polarisation, see
\citet{V15}). We considered prolate shaped large silicate grains with
axial ratio $a/b =2$. We made the further simplifying assumptions that
the largest radius of aligned silicates and the exponent of the size
distribution is the same as derived by fitting the extinction curve,
so that for the fit to the polarisation curves we were left with one
free parameter $r_{-}^{\rm {pol}}$.

We applied the dust model and fit the extinction and polarisation
curves with eight free parameters: $q$, $\Upsilon_{\rm {aC}}$,
$\Upsilon_{\rm {gr}}$, $\Upsilon_{\rm {PAH}}$, $\Upsilon_{\rm {sSi}}$,
$r_{-}^{\rm {pol}}$, $r_{+}^{\rm {Si}}$, and $r_{+}^{\rm
  {aC}}$. Best--fit parameters and their $1\,\sigma$ errors are
computed by a least-square technique utilizing the
Levenberg--Marquardt algorithm as implemented in
MPFIT \footnote{http://purl.com/net/mpfit} (\citet{Markwardt, More77,
  More93}). The algorithm is able to find the local minima, and to
identify the global minimum we need to apply it starting from many
different initial values. For each particle radius $100\,{\rm{nm}} <
r_{+}^{\rm{aC}} < 800$\,nm we computed the reduced $\chi^2$ of the fit
to the extinction $\chi_{\rm e}^2(r)$ and polarisation $\chi_{\rm
  p}^2(r)$ curve. The model that simultaneously fits both curves best
was selected from the minimum $\chi$ condition and where each of the
eight fitting parameters is given the same weight. We found that in our
model $\chi^2(r)$ was generally well behaved showing a (nearly)
parabolic shape and a global minimum.

%%%%%%%%%%%%%%%%%%%%

\begin{table*}[h!tb]

\begin{center}
  \caption {Fit parameters of the dust models.
\label{para.tab}}
  \small

 \begin{tabular}{c c r r c c l l l }
\hline \hline
Target & [Si]$^{\rm {tot}}$/[H] & [C]$^{\rm {tot}}$/[H] & [C]$^{\rm {gr}}$/[H] & [C]$^{\rm {PAH}}$/[H] & $q$ & $r^{\rm{pol}}_{-}$ & $r_{+}^{\rm {aC}}$ & $r_{+}^{\rm {Si}}$  \\
       & (ppm)& (ppm)& (ppm)& (ppm)& & (nm)& (nm)& (nm) \\
 & (1) & (2) & (3) & (4) & (5) & (6) & (7) & (8) \\
\hline 
HD149404&   21&$  102 \pm     4$&$   12$&$    8$&$  3.0$&$  101 \pm    23$&$  266 \pm    10$&$   348 \pm    13$ \\
HD151804&   20&$   88 \pm     2$&$    6$&$    6$&$  2.4$&$  146 \pm     9$&$  358 \pm    11$&$   234 \pm     7$ \\
HD152235&   24&$  113 \pm     3$&$   15$&$    7$&$  2.7$&$  101 \pm    17$&$  279 \pm     7$&$   194 \pm     5$ \\
HD152248&   19&$   74 \pm     2$&$    4$&$    7$&$  2.9$&$   66 \pm     6$&$  419 \pm    12$&$   253 \pm     7$ \\
HD152249&   19&$   74 \pm     3$&$    9$&$    6$&$  3.0$&$  218 \pm     9$&$  346 \pm    24$&$   263 \pm    18$ \\
HD152408&   21&$   79 \pm     2$&$    9$&$    6$&$  2.2$&$  217 \pm     9$&$  335 \pm     9$&$   232 \pm     7$ \\
HD153919&   21&$   77 \pm     3$&$   12$&$    6$&$  2.8$&$  104 \pm     2$&$  318 \pm    21$&$   324 \pm    21$ \\
\hline
\end{tabular}
\end{center} {NOTES --} We convert specific to absolute dust
abundances taking 15\,ppm of [Si]/[H] locked in large ($6\,{\rm {nm}}
\la r \la r_{+}^{\rm Si}$) and the rest of Si in small ($r < 6$\,nm)
silicates.  This gives absolute abundances of (1) all Si, (2) all C,
and of carbon in (3) graphite, and (4) PAH, respectively. Error of the
abundance is $1 \sigma \sim 1$\,ppm if not specified otherwise. (5)
Exponent $q$ of the dust size distribution, $n(r) \propto r^{-q}$
(\citet{Mathis77}), $1 \sigma \sim 0.02$. (6) Minimum radius of
aligned silicates ($r^{\rm{pol}}_{-}$). Maximum radius of (7)
silicates and (8) amorphous carbon grains.
\end{table*}

%%%%%%%%%%%%%%%%%%%%%%%%%%%%%%%%%%%%%%%%%%%%

\begin{figure} [h!tb]
  \begin{center}
\includegraphics[width=10cm,clip=true,trim=2.cm 4.5cm 1.cm 5.cm]{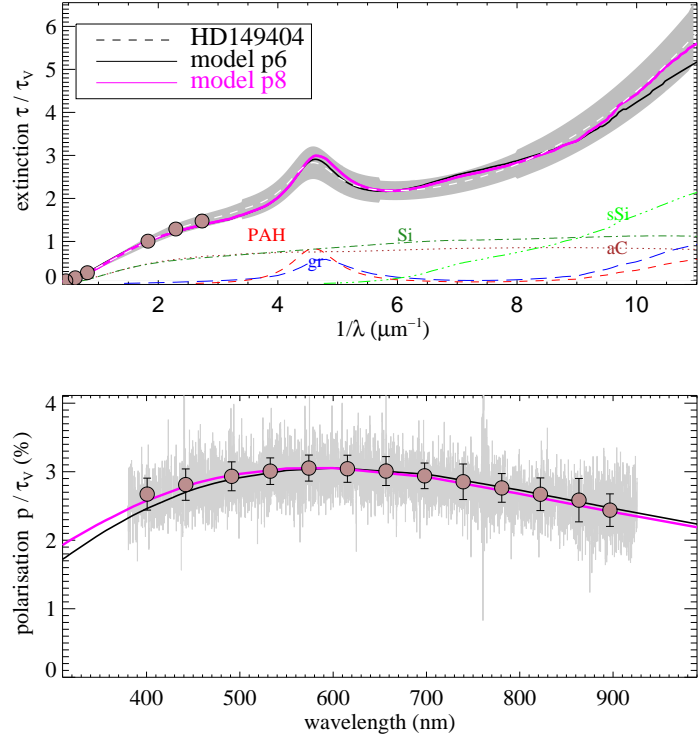}
\end{center}
\caption{Fit to the extinction (\citet{Valencic, Fitzpatrick04}) and
  polarisation (this work) curve of HD\,149404.  Model p8 (magenta)
  and p6 (black) are shown by full lines. {\it {Top:}} Observed
  extinction (white dashed line) with 1$\sigma$ uncertainty (hashed
  area in grey), symbols refer to the $UBVJHK$ photometry.  The
  contribution of the different dust populations to the extinction is
  given for model p8 as labeled. {\it {Bottom:}} FORS polarisation
  spectrum (grey) and reduced to a spectral resolution of
  $\lambda/\Delta \lambda \sim 50$ (filled circles) with 1\,$\sigma$
  error bars. \label{HD149404.fig}}
\end{figure}

\begin{figure} [h!tb]
\begin{center}
\includegraphics[width=10cm,clip=true,trim=2.cm 4.5cm 1.cm 5.cm]{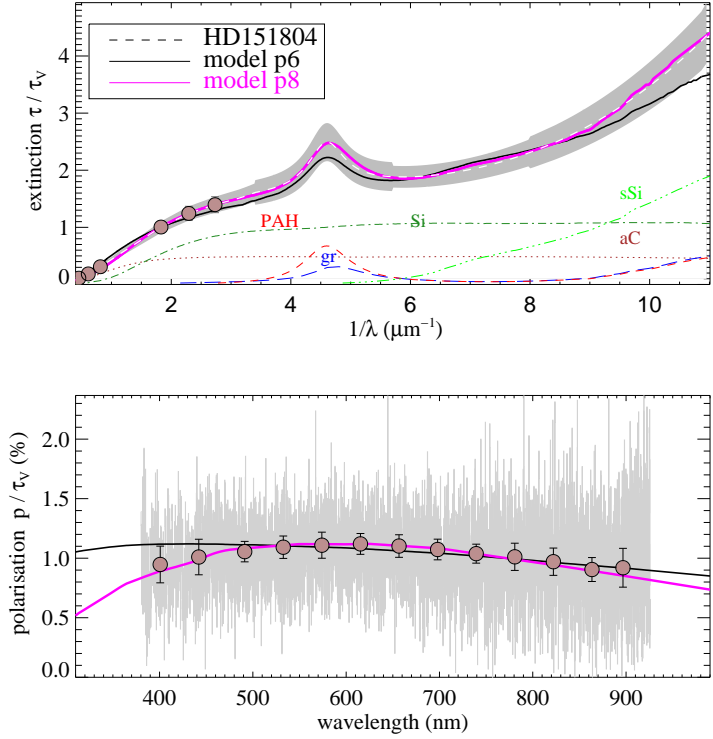}
\end{center}
\caption{Same as Fig.~\ref{HD149404.fig} for HD\,151804.}
\end{figure}

\begin{figure} [h!tb]
\begin{center}
\includegraphics[width=10cm,clip=true,trim=2.cm 4.5cm 1.cm 5.cm]{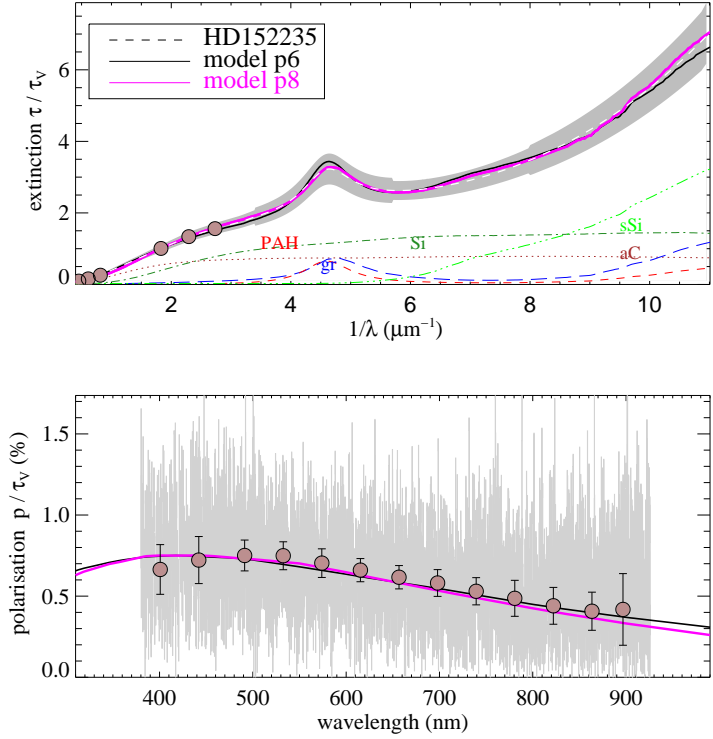}
\end{center}
\caption{Same as Fig.~\ref{HD149404.fig} for HD\,152235.}
\end{figure}

\begin{figure} [h!tb]
\begin{center}
\includegraphics[width=10cm,clip=true,trim=2.cm 4.5cm 1.cm 5.cm]{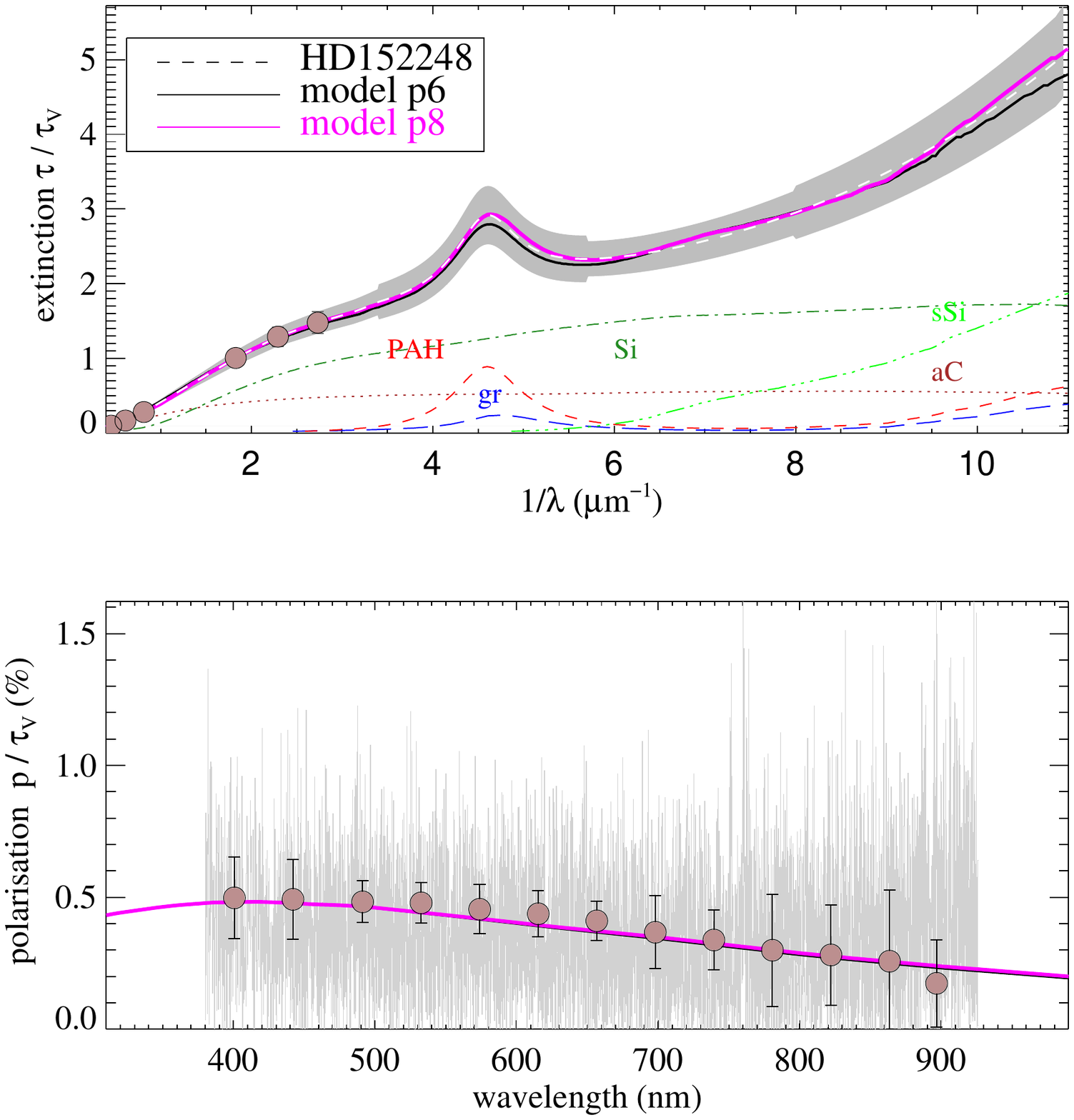}
\end{center}
\caption{Same as Fig.~\ref{HD149404.fig} for HD\,152248.}
\end{figure}

\begin{figure} [h!tb]
\begin{center}
\includegraphics[width=10cm,clip=true,trim=2.cm 4.5cm 1.cm 5.cm]{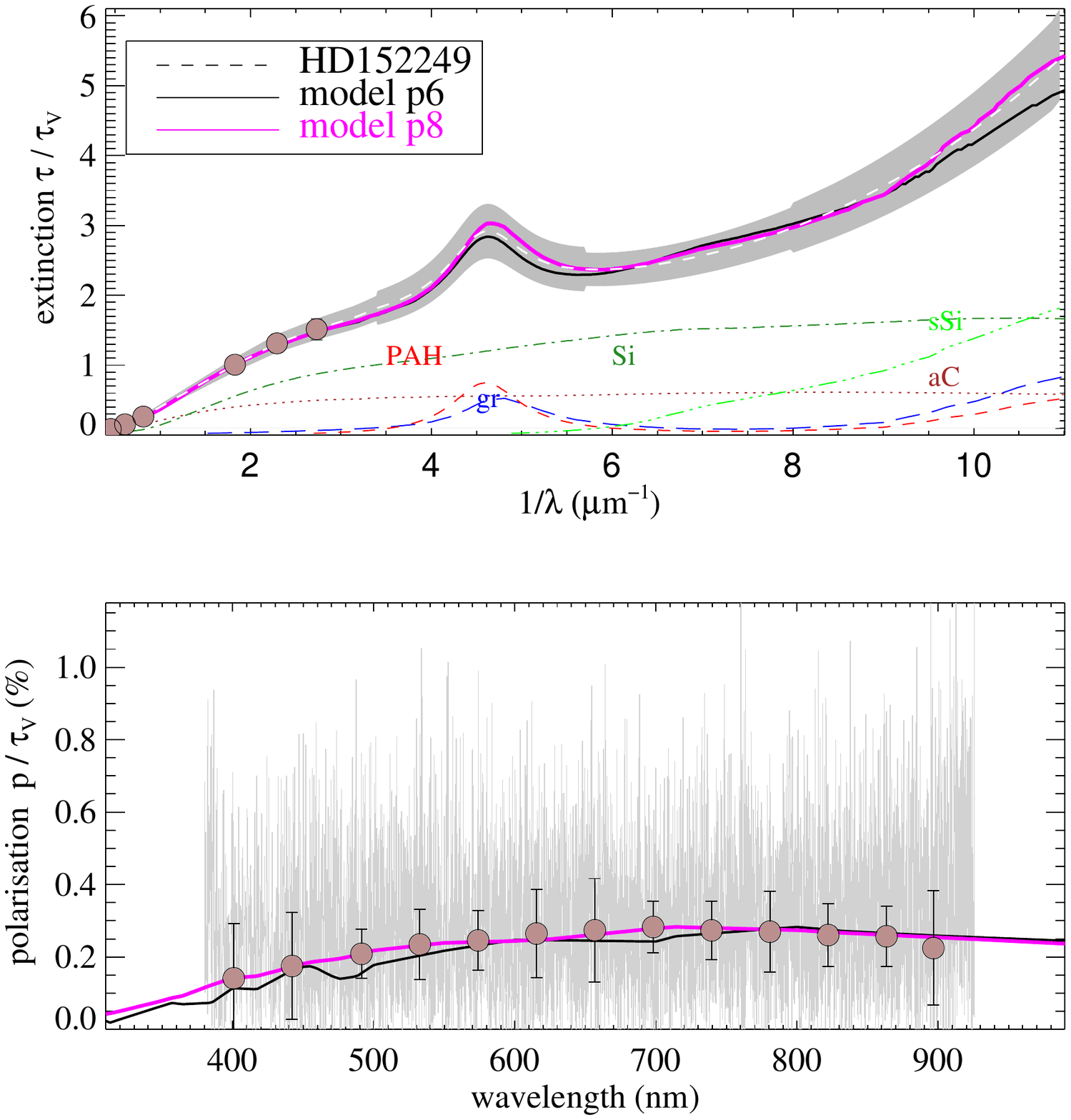}
\end{center}
\caption{Same as Fig.~\ref{HD149404.fig} for HD\,152249.}
\end{figure}

\begin{figure} [h!tb]
\begin{center}
\includegraphics[width=10cm,clip=true,trim=2.cm 4.5cm 1.cm 5.cm]{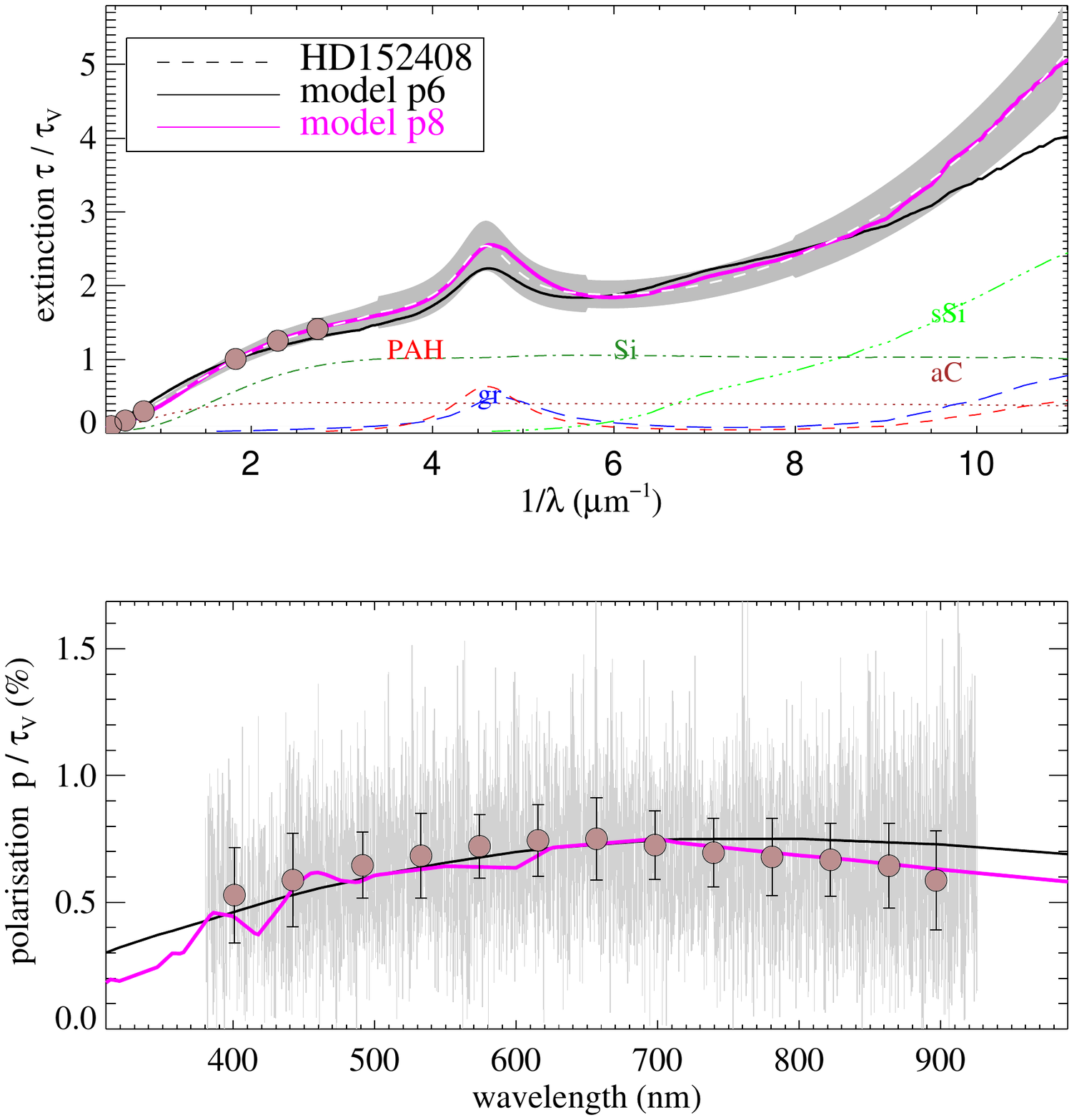}
\end{center}
\caption{Same as Fig.~\ref{HD149404.fig} for HD\,152408.}
\end{figure}

\begin{figure} [h!tb]
\begin{center}
\includegraphics[width=10cm,clip=true,trim=2.cm 4.5cm 1.cm 5.cm]{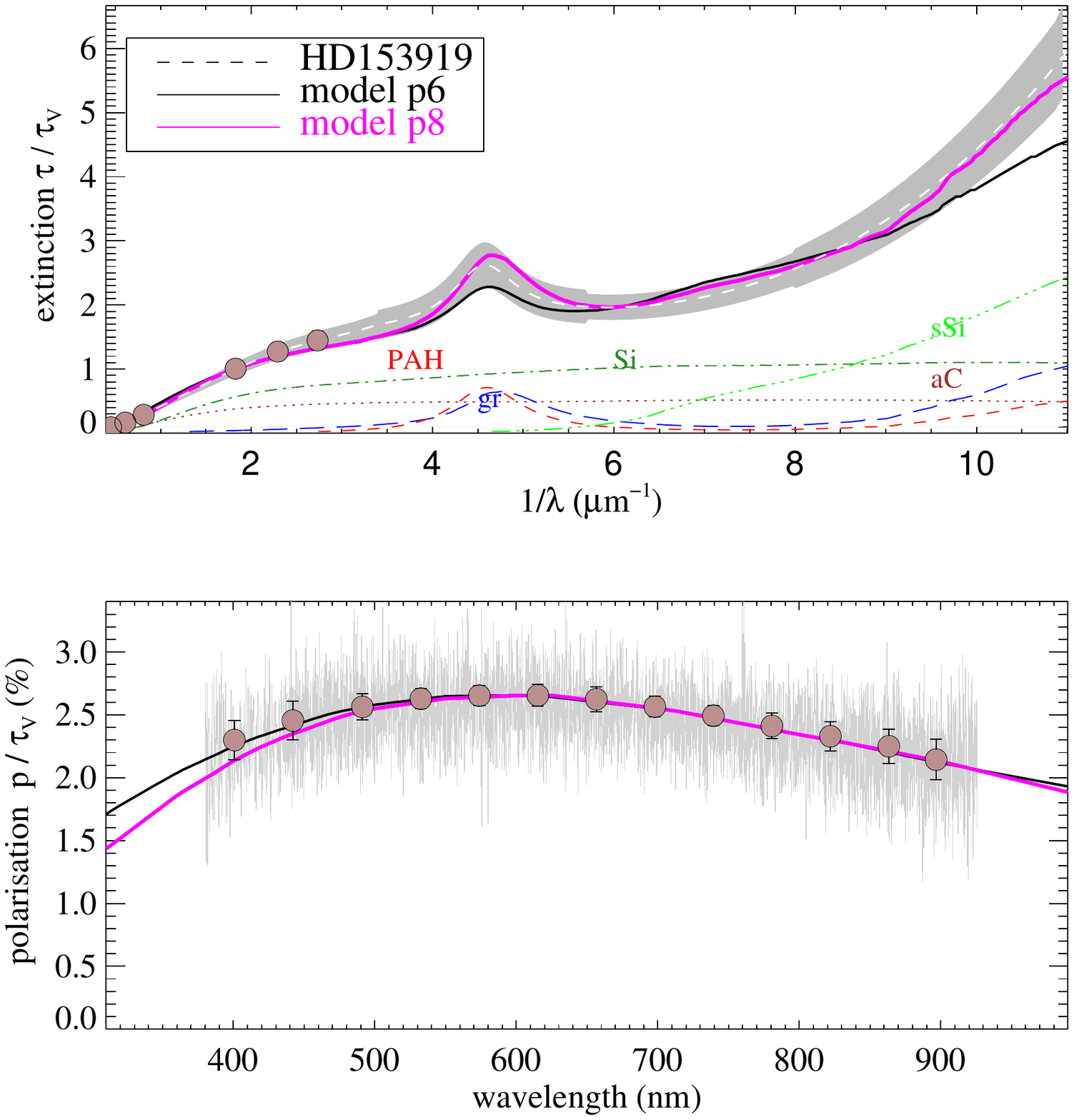}
\end{center}
\caption{Same as Fig.~\ref{HD149404.fig} for HD\,153919. \label{HD153919.fig}}
\end{figure}

%%%%%%%%%%%%%%%%%%%%%%%%%%%%%%%%%%%%%%%%%%%%%%%%

\section{Results} \label{result.sec}

In this Sect. we present our best-fits to the extinction and
polarisation spectra of the seven targets of Sect.~\ref{obs.sec}
obtained by applying the dust model and fitting procedure described in
Sect.~\ref{model.sec}. We will call the model with eight free
parameters 'p8'.  Figures~\ref{HD149404.fig} -- \ref{HD153919.fig}
show the data and best--fits for each target, and are organised as
follows. In the upper panel we display the extinction data
$\tau(\lambda^{-1})/\tau_{V}$ of the target with a white dashed line
on a grey background that represents the 1\,$\sigma$ uncertainty.  The
circles show the photometry in the $UBVJHK$ bands. Model p8 is shown
with a magenta line. For this model we show the contribution of each
dust population to the extinction. The lower panel in each figure
shows the linear polarisation spectrum $p(\lambda)/\tau_{V}$.  The
original (non rebinned) FORS spectra are represented with a grey line,
while the circles represent the values rebinned to a spectral
resolution $\Delta \lambda/\lambda \sim 50$.  The best-fit (obtained
with large aligned silicate grains) is represented with a magenta
line.

We applied the conversion of specific to absolute dust abundances as
described in Sect.~\ref{obs.sec}. The total Si and C abundance in the
dust model is denoted by [Si]$^{\rm {tot}}$/[H] and [C]$^{\rm
  {tot}}$/[H], respectively. They are given for each target in
Table~\ref{para.tab} together with absolute C abundance in graphite
and PAHs and the other fitting parameters. 

{{ Cosmic element abundances are difficult to estimate and even more
    so is their depletion in dust. The sun's heavy elements abundances
    (\citet{Asplund09}) are higher than for B star that are in the
    solar vicinity. Therefore, we take as reference a gas phase carbon
    abundance $214 \pm 20$\,ppm (\citet{Nieva, Lyubimkov,
      Snow}). \citet{Nieva} report a C abundance in ISM dust of $123
    \pm 23$\,ppm that can be reduced to $84 \pm 23$\,ppm when the
    higher gas phase abundance by \citet{Sofia} is
    considered. \citet{Parvathi} gives dust phase abundances of C for
    11 sight-lines at a $\ge 3\sigma$ confidence level. We correct
    their values to the above reference and derive a median
    [C]/[H]$_{\rm {dust}} = 102 \pm 47$\,ppm. \citet{Gerin} finds a
    gas phase C abundance between 50 - 240\,ppm.  If one assumes that
    50\% of C is depleted into dust grains the C abundance in the dust
    is 75\,ppm. { {The interstellar absorption feature at
        3.4\,$\mu$m can be explained by hydrogenated amorphous carbon
        with a required number of carbon atoms in these grains of 72
        -- 97\,ppm derived by \citet{Duley98} and 80\,ppm by
        \citet{Furton99}.}}  In model p8 the median C abundance in
    dust is $79 \pm 14$\,ppm, which is in agreement with the previous
    estimates.

    \citet{Nieva} derive [Si]/[H]$_{\rm {dust}} = 29.4 \pm 3.6$\,ppm,
    whereas \citet{VH10} find, for similar sight-lines in the diffuse
    ISM as in our sample, $15 \la \rm{[Si]/[H]}_{\rm {dust}} \la
    31.4$\,(ppm) and a mean of $24 \pm 3$\,ppm.  In model p8 the Si
    dust abundance range between 19 - 24\,ppm at a mean of $21 \pm
    2$\,ppm. However, there are also open discussions on the Fe and O
    dust phase abundances (\citet{Dwek16, Jenkins}) and the detailed
    mineralogy of silicate grains (\citet{H10}). }}

The polarisation curves are explained by large aligned silicates with
{{average minimum alignment radii of $<r_{-}^{\rm {pol}}> = 140
    \pm 56$\,(nm) and maximum radius $<r_{+}^{\rm {Si}}> = 264 \pm
    50$\,(nm), respectively. For HD\,152248 and HD\,152249 the
    observed polarisation is {{below 0.5\%}}.  The upper radius of
    carbon grains is $\sim 335 \pm 50$\,nm, so about 25\,\%}} larger
than for silicate grains (Table~\ref{para.tab}). Probably, this is
because the efficiency of destruction of carbonaceous particles is
lower (\citet{Slavin15}). Note that \citet{HV14} presented a dust
model fitting extinction data by grain growth. They need to tune the
accretion and coagulation process and increase the sticking
coefficient of silicates so that silicates become larger than carbon
grains.

We have extensively explored whether other dust models with a smaller
number of free parameters could satisfy extinction, polarisation and
abundance constraints. As a first attempt, we simplified our p8 model
by assuming that the upper radius of large silicates and carbon dust
are identical ($r_{+}^{\rm {Si}} = r_{+}^{\rm {aC}})$. The best-fits
to the extinction and polarisation curve of such a model are good.
{ {However, it requires more C than model p8.}}

\begin{figure} [h!tb]
%\centering
\includegraphics[width=10cm,clip=true,trim=1.cm 1.cm 0.cm 1.cm]{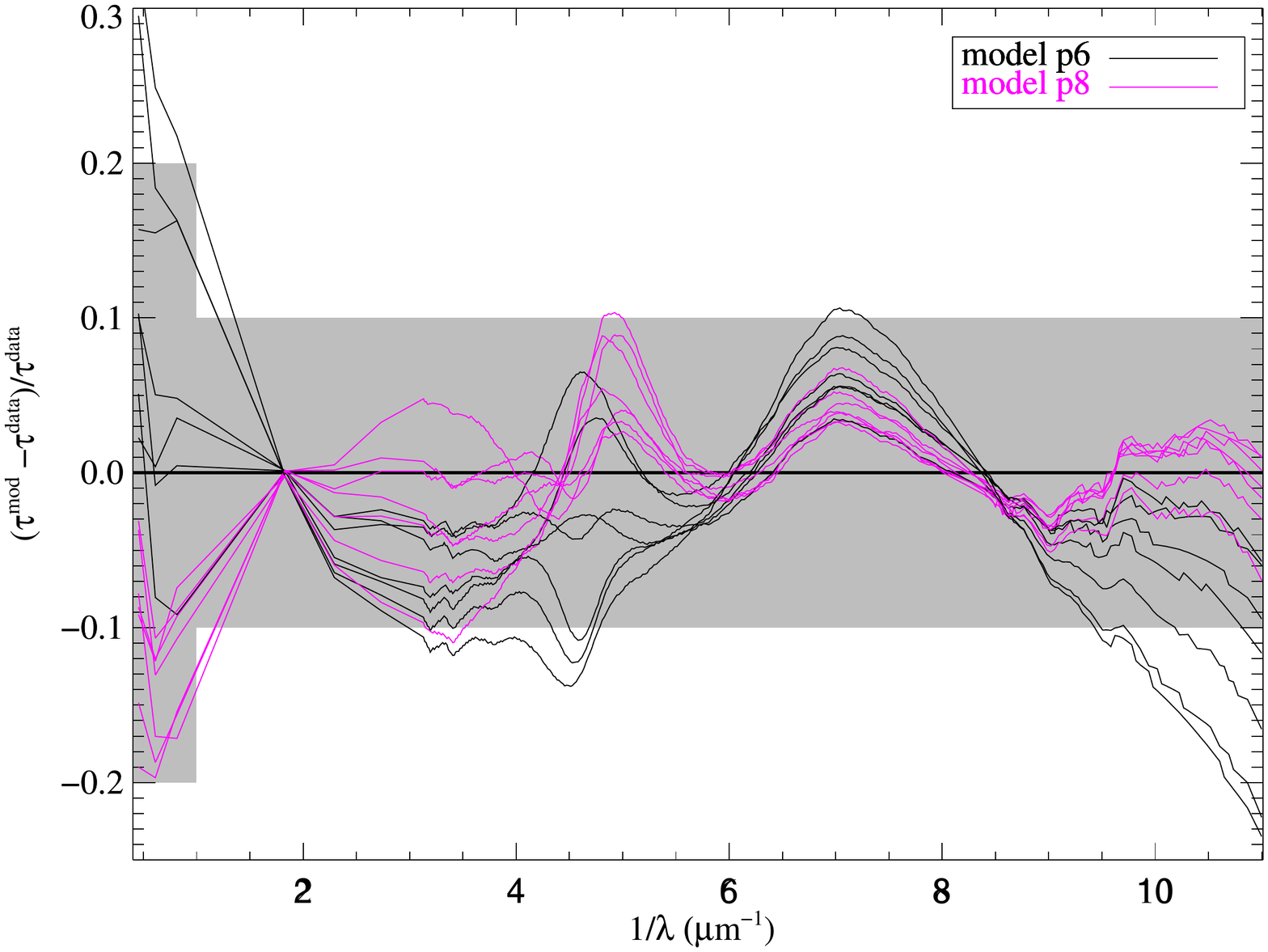}
\includegraphics[width=10cm,clip=true,trim=1.cm 1.cm 0.cm 1.cm]{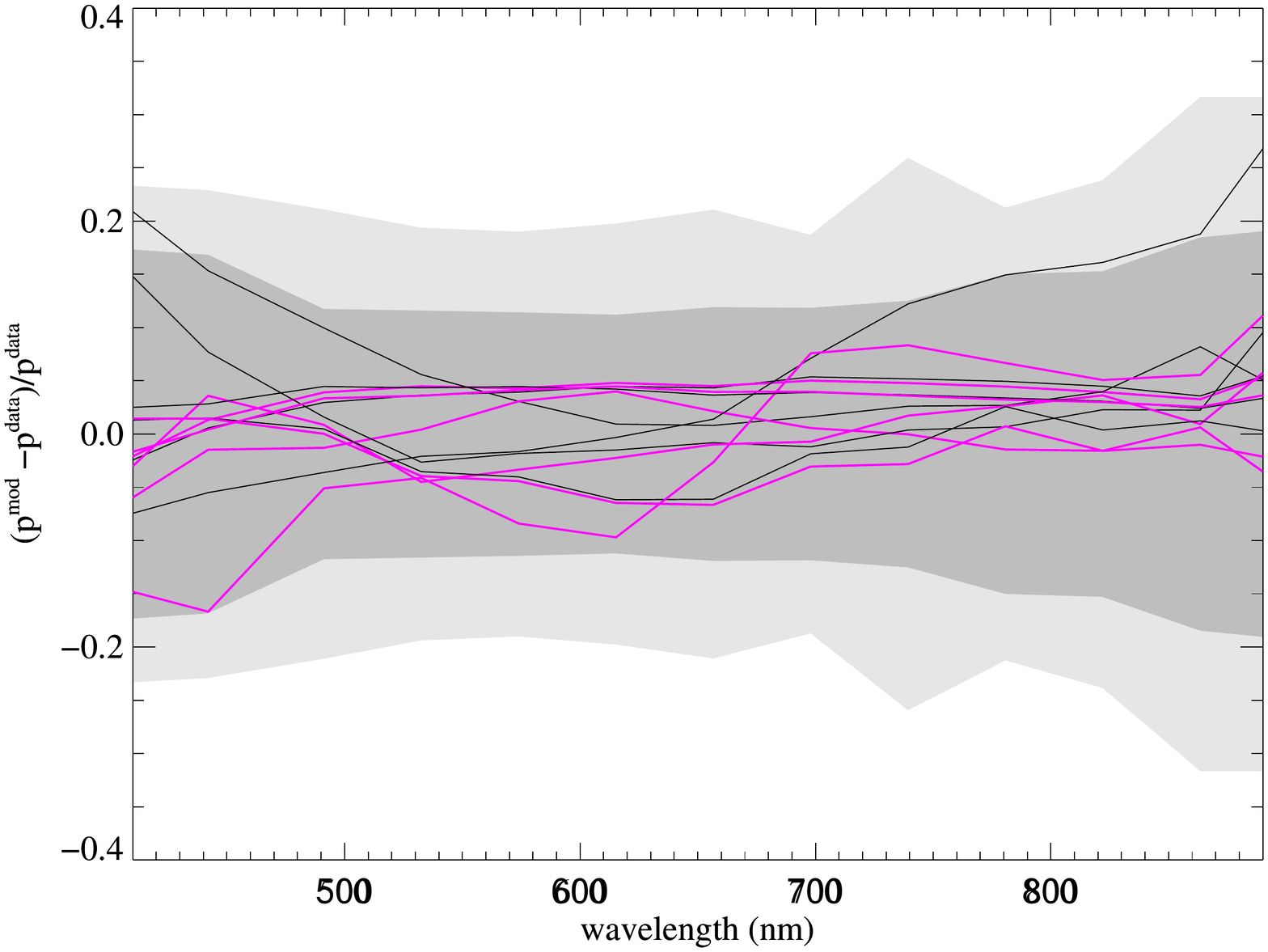}
\caption{Residuals of model p8 (magenta) and model p6 (black) to the
  extinction (top) and polarisation data (bottom) presented for
  targets in Fig.~\ref{HD149404.fig} -- Fig.~\ref{HD153919.fig}. The
hashed area in grey indicates a typical 1$\sigma$ error and the light
grey area in the bottom panel mark the maximum 1$\sigma$ error in the
binned polarisation spectra. \label{residu.fig}}
\end{figure}

We considered also a model with six free parameters (that we will call
p6), by fixing $q=3.1$, assuming that the C abundance in PAHs is 8\,\%
the C abundance in large carbon grains ($\Upsilon_{\rm {PAH}} = 8\,\%
\Upsilon_{\rm {aC}}$), and keeping as free parameters the quantities
$\Upsilon_{\rm {aC}}$, $\Upsilon_{\rm {gr}}$, $\Upsilon_{\rm {sSi}}$,
$r_{-}^{\rm {pol}}$, $r_{+}^{\rm {Si}}$, and $r_{+}^{\rm {aC}}$.  It
is shown as black lines in Figs.~\ref{HD149404.fig} --
\ref{HD153919.fig}. Model p6 provides good fits to the extinction and
polarisation spectra, however it produces somewhat larger deviations from the
data than model p8.  This is evident inspecting residuals of fits
to the extinction, $(\tau^{\rm{mod}} - \tau^{\rm {data}})/\tau^{\rm
  {data}}$, and polarisation, $(p^{\rm{mod}} - p^{\rm {data}})/p^{\rm
  {data}}$. They are displayed in Fig.~\ref{residu.fig}. For some
stars model p6 has difficulties accounting for the steep rise in the
far UV (at $x \simgreat 10\,\mu$m$^{-1}$), and for two stars it
underestimates the extinction bump, and overestimates the near IR
extinction. Model p6 accounts for the [Si]$^{\rm {tot}}$/[H] abundance
constraints but requires 50\,\% more C than {{model p8}}.

Finally, we apply a model where we include contributions by large
grains and PAHs and assume $r_{+}^{\rm {Si}} = r_{+}^{\rm {aC}}$. This
model ignores very small silicates and graphite grains.  It reproduces
the polarisation spectra. The model fits the extinction curve for $x
\simless 7 \mu$m$^{-1}$. However, at shorter wavelength it stays
rather flat and fails to account for the steep extinction rise in the
far UV.  This model accounts for the [Si]$^{\rm {tot}}$/[H] abundance
constraints and requires 30\,\% more C than {{model p8}}.

%\begin{figure} [h!tb]
%%\centering
%\includegraphics[width=8.6cm,clip=true,trim=4.3cm 9.5cm 3.7cm 10cm]{./Fig/pl_extin.pdf}
%\caption{Median of the best--fit extinction curves normalized to the $K$
%  band $\tau/\tau_{K}$ for models nvsg (cyan), p5 (magenta) and p7
%  (black).  Dots show extinction curves as in Figs.~\ref{HD149404.fig}
%  -- \ref{HD153919.fig} and  same colour code. \label{extin.fig}}
%\end{figure}

%%%%%%%%%%%%%%%%%%%%%%%%%%%%%%%%%%%%%%%%%%%%%%%%%%%%%%%%%

\section{Conclusion} \label{conclusion.sec}

We have presented some preliminary results of our Large Interstellar
Polarisation Survey (LIPS) at the VLT, using a selected sub-sample of
seven stars along lines-of-sight towards the Sco OB1 association. For
these stars we have obtained high S/N polarisation spectra with the
FORS2 instrument in the spectral range 365 -- 920\,nm, while
extinction curves were taken from the literature \citep{Valencic}. The
mean $R_{V} \sim 3.7$ of our sample is somewhat larger than the
typical value of $R_{V} \sim 3.1$ of the diffuse ISM (\citet{FM07}).

We explained the dust extinction and polarisation spectra with a model of the
interstellar dust that includes a population of carbon and silicate
grains with a power-law size distribution ranging from the molecular
domain (0.5\,nm and PAHs) up to 800\,nm. Particles with radii larger
than 6\,nm are spheroids. Dust cross section computed in the
optical/UV with different sets of optical constants (\citet{Zubko,
  Draine03, Jones12, Jones13}) are similar.

Extinction and polarisation data are best-fit by assuming dust models
with different numbers of free parameters. We vary the exponent of the
size distribution $q$, specific abundances of C in amorphous carbon
$\Upsilon_{\rm {aC}}$, graphite $\Upsilon_{\rm {gr}}$, and PAHs
$\Upsilon_{\rm {PAH}}$, as well as Si in small silicates
$\Upsilon_{\rm {sSi}}$, the minimum radius of aligned silicates
$r_{-}^{\rm {pol}}$, and the upper grain radius of silicates
$r_{+}^{\rm {Si}}$ and amorphous carbon $r_{+}^{\rm {aC}}$.  A model
where all eight parameters are varied {{seems to be in reasonable}}
  agreement with present abundance constraints of C and Si in
  dust. Equally good fits are found assuming the same upper grain
  radius for both dust components. However, such a model overestimates
  the C abundance in two out of seven sight lines studied. We find
  another model with less (6) free parameters that fit data down to
  the lowest IUE observing range ($x \simless 8.6
  \mu$m$^{-1}$). However, it overestimates the C abundance in dust by
  $\sim 50$ \%. We analysed another model that includes large grains
  and a PAH component but ignores small graphite and small silicates.
  We keep in that model $r_{+}^{\rm {Si}} = r_{+}^{\rm {aC}}$. Such a
  model provides fits at somewhat larger reduced $\chi^2$, {
    {requires $\sim 30\%$ more carbon than model p8, and}} fails
    accounting for the steep rise of the extinction curve in the far
    UV at $x \simgreat 7 \mu$m$^{-1}$. We take it as a demonstration
    that beside PAHs, dust populations of very small silicate and
    graphite grains are needed.

    We conclude that {{data are}} best fit by a model with eight
    free parameters. From our modelling of the sight lines we found
    that the exponent of the dust size distribution is $q \sim 2.7 \pm
    0.3$.  The upper size of large silicates ranges between 194 and
    348\,nm, and that of amorphous carbon grains is $\sim 25\,\%$
    larger. This is in agreement with recent estimates of grain
    destruction efficiencies (\citet{Slavin15}).  We have converted
    the specific abundances of the model into absolute abundances by
    assuming that 15\,ppm of Si is locked in large silicate
    grains. Our model requires $6 \pm 2$\,ppm of Si in very small
    silicates. The abundance of C in graphite compared to the total C
    in dust is $\Upsilon_{\rm {gr}} / \Upsilon_{\rm {tot}} = 11 \pm
    3$\,(\%) and that of PAH $\Upsilon_{\rm {PAH}} / \Upsilon_{\rm
      {tot}} = 8 \pm 1$\,(\%). The polarisation curves are explained
    by large aligned silicates with minimum alignment radii $66 \la
    r_{-}^{\rm {pol}} \la 218$\,(nm).  Typical numbers of the
    individual dust abundances and other parameters derived from the
    complete LIPS sample will be presented in a forthcoming paper.

\section{Acknowledgments}

{ {The anonymous referees are thanked for their careful read and
    helpful suggestions improving the quality of the article. In
    particular Referee 2 is warmly thanked for drawing our attention
    to cosmic abundance constraints in the gas phase and solids of the
    ISM, and also the papers discussing the bulk density of amorphous
    carbon.  We thank Endrik Krügel for various discussions.}} NVV was
partly supported by the RFBR grant 16-02-00194. This work is based on
observations made with ESO Telescopes at the La Silla Paranal
Observatory under programme ID 095.C-0855(A) and 096.C-0159(A).  This
research has made use of the SIMBAD database, operated at the CDS,
Strasbourg, France.

\end{document}